\documentstyle[12pt]{article}
\input{tcilatex}

\begin{document}

\title{A Quantum Particle Undergoing Continuous Observation}
\author{V.P. Belavkin\thanks{%
M.I.E.M., D. Vusovsky 3/12, Moscow 109028, USSR} \ and P. Staszewski\thanks{%
Institute of Physics, N. Copernicus University, Toru\'n, Poland}}
\date{December 1988\\
Published in: {\it Physics Letters A}, {\bf 140} (1989) No 7,8, pp 359 --362 
}
\maketitle

\begin{abstract}
A stochastic model for the continuous nondemolition ohservation of the
position of a quantum particle in a potential field and a boson reservoir is
given. lt is shown that any Gaussian wave function evolving according to the
posterior wave equation with a quadratic potential collapses to a Gaussian
wave packet given by the stationary solution of this equation..
\end{abstract}

The recently developed methods of quantum stochastic calculus \cite%
{bib:12,bib:bquanpar2} can serve for the description of the time-development
of continuously observed quantum systems \cite{bib:bel1,bib:9,bib:grand4}.

We apply this approach to describe the time-behaviour of a
one-di\-men\-sio\-nal quantum particle in the field of the linear force $%
F=kx+mg$. The effect of coupling of the particle to a measuring apparatus is
represented by an extra stochastic force. Assuming a non-ideal indirect
observation of the particle position one can select such an observation
channel that the observation is nondemolition \cite{bib:9,bib:grand4}, i.e.
it does not affect the actual as well as the future states of the perturbed
particle. The posterior dynamics of the observed particle is then given by
the nonlinear stochastic wave equations rigorously derived by quantum
filtering method in \cite{bib:9,bib:ref17}.

It is shown that the Gaussian wave packet evolves to the asymptotic
stationary one the width of which (identified with the standard deviation)
in the coordinate representation is given by the formula $\tau _{q}=\hbar
/2m[(\kappa ^{2}+\lambda ^{2})^{1/2}-\kappa ]^{1/4}$, where $\kappa =k/\hbar 
$ and $\lambda $ is the accuracy coefficient of the nondemolition
measurement of the particle position. This phenomenon which cannot be
explained by the Schr\"{o}dinger equation (describing the time-evolution of
the unobserved quantum system) belongs to the class of phenomena called
watchdog effects \cite{bib:bzeno5}.

Let us assume that the measuring apparatus is modelled by the Bose field.
The motion of the particle in a potential $\phi$ is distorted by the
apparatus so the time-derivative $\dot P$ of the momentum of the particle is
no longer equal to $F(X)=-\phi^{\prime}(X)$ ($\phi^{\prime}=\partial\phi/%
\partial x$). The Heisenberg equations describing the time-development of
the momentum and the position of the observed particle have the form \cite%
{bib:b2} 
\begin{equation}
\dot P(t)-F\bigl(X(t)\bigr)=f(t),\quad \dot X(t)=\frac 1m P(t),  \label{eq:1}
\end{equation}
where the force $f(t)$ in our model is taken as 
\begin{equation}
f(t)=\frac\hbar i(\lambda/2)^{1/2}[a^\dagger (t)-a(t)]=
(2\lambda)^{1/2}\hbar\Im a^\dagger (t),\quad \lambda>0.  \label{eq:2}
\end{equation}
Here $a(t)=b_0(t)$, $a^\dagger(t)=b^\dagger_0(t)$ are the annihilation and
creation quantum noise operators with the canonical commutation relations 
\begin{equation}
[a(t),a^\dagger (t)]=\delta(t- t^{\prime}),\quad [a(t),a(t^{\prime})]=0\,.
\label{eq:3}
\end{equation}
given by the standard boson field operators $b_s$, $b_s^\dagger$, on the
half of line $s\leq0$ with free evolution $b_s(t)=b_{s-ct}$, $%
b_s^\dagger=b^\dagger_{s-ct}$ at $s=0$.

The position of the particle is assumed to be observed indirectly, together
with some noise $e(t)$ (error), therefore the measured quantity is 
\begin{equation}
y(t)=(2\lambda )^{1/2}X(t)+e(t).  \label{eq:4}
\end{equation}
it is easy to check with the help of (\ref{eq:3}) that if 
\begin{equation}
e(t)=a(t)+a^{\dagger }(t)=2\Re a(t)  \label{eq:5}
\end{equation}
then for $P(t)$ and $X(t)$ satisfying (\ref{eq:1}) the following commutation
relations hold, 
\begin{equation}
\lbrack P(t),y(t^{\prime })]=0\,,\quad \lbrack X(t),y(t^{\prime
})]=0\,,\quad \forall \,t^{\prime }\leq t.  \label{eq:6}
\end{equation}
From these relations it follows that the preparation for the measurement of
any functional $Y$ of the past operators $y(t^{\prime })$, $t^{\prime }\leq
t $ does not affect $P(t)$ and $X(t)$ as well as any other Heisenberg
operator $Z(t^{\prime })$ of the particle at $t^{\prime }\geq t$. In other
words, the condition (\ref{eq:6}) means that the measurement of $Y$ disturbs 
{\em a priori\/} neither the present nor the future state of the particle.

Note, that $y(t)=y^{\dagger }(t)$ and 
\begin{equation}
\lbrack y(t),y(t^{\prime })]=0\,,\quad \forall \,t,t^{\prime }\,,
\label{eq:7}
\end{equation}
therefore $y(t)$ can be continuously measured in time as a classical
quantity. Following refs. \cite{bib:9,bib:ref11}, one can say that the
continual measurement of $y$ is nondemolition with respect to the
time-evolution of the system.

Obviously, the minimal distortion of motion (\ref{eq:1}) will be obtained
for the Bose field in a vacuum state. In such a situation $e$ and $f$ are
white noises, $e$ has standard intensity $1$, while the intensity of $f$ is
proportional to the measurement accuracy: 
\begin{eqnarray}
\langle f(t)\rangle =\langle e(t)\rangle =0\,, &\mbox{}&\,\langle
f(t)f(t^{\prime })\rangle =\frac{1}{2}\lambda \hbar ^{2}\delta (t-t^{\prime
})\,,  \nonumber \\
\langle e(t)e(t^{\prime })\rangle &=&\delta (t-t^{\prime })\,.  \label{eq:8}
\end{eqnarray}
It is convenient to rewrite the equations of motion (\ref{eq:1}) in the form
of the Ito quantum stochastic differential equations \cite%
{bib:12,bib:9,bib:grand4,bib:b2}. We obtain 
\begin{eqnarray}
{\rm d}P(t) &=&\frac{i}{\hbar }[\phi \Big(X(t)\Big),P(t)]{\rm d}t+(\lambda
/2)^{1/2}[X(t),P(t)][{\rm d}A(t)-{\rm d}A^{\dagger }(t)]\,,  \nonumber \\
{\rm d}X(t) &=&\frac{i}{\hbar }\,[P^{2}(t)/2m,X(t)]{\rm d}t\,,  \label{eq:9}
\end{eqnarray}
where ${\rm d}A(t)=A(t+{\rm d}t)-A(t)$ is the stochastic differential of the
standard quantum Brownian motion, $A(t)|_{t=0}=0$, with the generalized
derivative $a(t)={\rm d}A(t)/{\rm d}t$. Using the quantum Ito formula \cite%
{bib:12} one can obtain from eqs. (\ref{eq:9}) the time-evolution equation
for any observable $Z$ (a self-adjoint polynom in $X$ and $P$) 
\begin{eqnarray}
{\rm d}Z(t) &=&\{-(i/\hbar )[Z(t),P^{2}(t)/2m+\phi \Big(X(t)\Big)]-\frac{1}{4%
}\,\lambda \lbrack X(t),[X(t),Z(t)]]\}{\rm d}t  \nonumber \\
&+&2\Re \,\{(\lambda /2)^{1/2}[X(t),Z(t)]{\rm d}A\}\,.  \label{eq:10}
\end{eqnarray}
Eq. (\ref{eq:10}) describes the prior stochastic dynamics of the particle
considered as an open quantum system; the term ``prior dynamics'' means that
the process $Z(t)$ is not conditioned by the results of the observation.
Obviously, the prior state of the particle in the Schr\"{o}dinger picture is
described by a mixed density matrix even if the initial state is pure (given
by the particle wave function $\Psi $ and the Fock vacuum vector).

The Bose field does not only disturb the system but also conveys some
information about it. This information is contained in the ``output field'' $%
a(t)_{{\rm out}}$ --- the field after interaction with the system in
question. The measured quantity $y(t)$ appearing in (\ref{eq:4}) can be
interpreted as $2\Re \,a(t)_{{\rm out}}$ while $e(t)=2\Re \,a(t)$ with $a(t)$
being the input field \cite{bib:b2}. Let us rewrite (\ref{eq:4}) in the form
of Ito quantum stochastic differential 
\begin{equation}
{\rm d}Y(t)=(2\lambda )^{1/2}X(t){\rm d}t+2\Re \,{\rm d}A(t)\,.
\label{eq:11}
\end{equation}
for the integral $Y(t)=\int_{0}^{t}y(t^{\prime }){\rm d}t^{\prime }$.

The posterior mean values $\hat{z}(t)$, i.e. the mean values of the process $%
Z(t)$ which is partially observed by means of $Y(t)$, are defined as
conditional expectations $\hat{z}(t)=\epsilon ^{t}(Z(t))$ with respect to
the observables $Y^{t}=\{Y(s)|s\leq t$. For a given $Z$, $\hat{z}$ is a
non-anticipating functional of the observed trajectories $q=\{q(t)\}$ of the
output process $Y$. If the initial state of the system is pure and the
initial state of the bath is a vacuum state then the posterior expectation
values are realized with the help of the stochastic wave function called the
posterior wave function. The posterior state is therefore a pure one \cite%
{bib:grand4}. According to refs. \cite{bib:9,bib:grand4}, the posterior
stochastic wave function $\hat{\varphi}(t,x)$ satisfies a new nonlinear
stochastic (posterior) wave equation which in our case has the form 
\begin{equation}
{\rm d}\hat{\varphi}+\left[ {\frac{i\hbar }{2m}}\,\hat{\varphi}^{\prime
\prime }+\left( {\frac{i}{\hbar }}\,\phi +{\frac{\lambda }{4}}\,(x-\hat{q}%
)^{2}\right) \hat{\varphi}\right] {\rm d}t=(\lambda /2)^{1/2}(x-\hat{q})x%
\hat{\varphi}{\rm d}\tilde{Y}\,,  \label{eq:12}
\end{equation}
with the initial condition $\hat{\varphi}(0,x)=\psi (x)$. In the above
formula 
\begin{equation}
\hat{q}(t)=\int \hat{\varphi}^{\ast }(t,x)x\hat{\varphi}(t,x){\rm d}x
\label{eq:13}
\end{equation}
and 
\begin{equation}
{\rm d}\tilde{Y}(t)={\rm d}Y(t)-(2\lambda )^{1/2}\hat{q}(t){\rm d}t
\label{eq:14}
\end{equation}
denotes the ito differential of the observed commutative innovating process,
which is equivalent to the standard Wiener one as $\langle {\rm d}\tilde{Y}%
(t)\rangle =0$ and $\Big({\rm d}\tilde{Y}(t)\Big)^{2}={\rm d}t$.

Let us now discuss the problem of the time-development of the posterior wave
function at $t\rightarrow\infty$. We shall assume that the initial state is
the Gaussian wave packet, 
\begin{equation}
\psi(x)=(2\pi\sigma^2_q)^{1/4}\exp\biggl(-{\frac{1}{4\sigma^2_q}}\, (x-q)^2+{%
\frac{i}{\hbar}}\,px\biggr)\,,  \label{eq:15}
\end{equation}
where $p$ and $q$ denote initial mean values of position and momentum of the
particle and $\sigma^2_q$ stands for the initial dispersion of the wave
packet in the coordinate representation. We shall prove that the solution $%
\hat\varphi(t,x)$ of eq. (\ref{eq:12}) corresponding to the initial
condition (\ref{eq:15}) has the form of a Gaussian packet, 
\begin{equation}
\hat\varphi(t,x)=c(t)\exp\{(1/\hbar)[m\omega(t)[x-\hat q(t)]^2/2+ i\hat
p(t)x]\}\,,  \label{eq:16}
\end{equation}
with the posterior mean values $\hat q(t)$ (given by (\ref{eq:13})) and 
\begin{equation}
\hat p(t)={\frac{\hbar}{i}}\int\hat\varphi^*(t,x)\hat\varphi^{\prime}(t,x)%
{\rm d} x  \label{eq:17}
\end{equation}
fulfilling linear filtration equations and $\omega(t)$ satisfying the
Riccati differential equation. The normalization factor $c(t)=
(2\pi\tau^2_q)^{-1/4}$ up to an inessential phase multiplier and $\tau^2_q=%
\widehat{q^2}-\hat q^2$ is a posterior dispersion of position.

For this purpose it is convenient to rewrite eq. (\ref{eq:12}) in terms of
complex osmotic velocity. Let us first introduce 
\begin{equation}
T(t,x)=R(t,x)+iS(t,x)=\hbar \ln \hat{\varphi}(t,x)\,.  \label{eq:18}
\end{equation}
Then by Ito's rule 
\[
{\rm d}G(\hat{\varphi})=G^{\prime }(\hat{\varphi}){\rm d}\hat{\varphi}+\frac{%
1}{2}\,G^{\prime \prime }(\hat{\varphi})({\rm d}\hat{\varphi})^{2} 
\]
applied to the function $G=\hbar \ln \hat{\varphi}$ and taking into account
that $({\rm d}\hat{\varphi})^{2}=\frac{1}{2}\,\lambda (x-\hat{q})^{2}\hat{%
\varphi}^{2}{\rm d}t$ we obtain eq. (\ref{eq:12}) in the from 
\begin{equation}
{\rm d}T+[\frac{1}{2}\,\hbar \lambda (x-\hat{q})^{2}+i\phi -(i/2m)(T^{\prime
2}+\hbar T^{\prime \prime })]{\rm d}t=(\lambda /2)^{1/2}\hbar (x-\hat{q})%
{\rm d}\tilde{Y}\,.  \label{eq:19}
\end{equation}
In terms of complex osmotic velocity 
\begin{equation}
W(t,x)=\frac{1}{m}\,T^{\prime }(t,x)=U(t,x)+iV(t,x)  \label{eq:20}
\end{equation}
eq. (\ref{eq:19}) can be rewritten as 
\begin{equation}
{\rm d}W+[(\hbar \lambda /m)(x-\hat{q})+(i/\hbar )\phi ^{\prime
}-i(WW^{\prime }+\hbar W^{\prime \prime }/2m)]{\rm d}t=(\lambda /2)^{1/2}{%
\frac{\hbar }{m}}\,{\rm d}\tilde{Y}\,.  \label{eq:21}
\end{equation}
We look for the solution of (\ref{eq:21}) corresponding to the linear force $%
F(x)=\hbar \kappa x+mg$ and to the initial condition 
\begin{equation}
W(0,x)={\frac{\hbar }{m}}\,{\frac{\Psi ^{\prime }(x)}{\Psi (x)}}={\frac{%
\hbar }{2m\sigma _{q}^{2}}}\,(x-q)+{\frac{i}{m}}\,p  \label{eq:22}
\end{equation}
in the linear form 
\begin{equation}
W(t,x)=\hat{w}(t)+\omega (t)x\,,  \label{eq:23}
\end{equation}
where in accordance with (\ref{eq:16}) 
\begin{equation}
\hat{w}=-\omega \hat{q}+{\frac{i}{m}}\,\hat{p}\,.  \label{eq:24}
\end{equation}
By inserting $W^{\prime }=\omega $, $W^{\prime \prime }=0$ into (\ref{eq:21}%
) we obtain the following system of equations for the coefficients $\hat{w}%
(t)=W(t,0)$ and $\omega (t)=W^{\prime }(t,0)$, 
\begin{equation}
{\rm d}\hat{w}(t)-i[g+\omega (t)\hat{w}(t)]{\rm d}t=(\lambda /2)^{1/2}{\frac{%
\lambda }{m}}\,{\rm d}\tilde{Y}(t)\,,  \label{eq:25}
\end{equation}
with $\hat{w}(0)={\frac{\hbar }{2m\sigma _{q}^{2}}}\,q+{\frac{i}{\hbar }}\,p$%
, 
\begin{equation}
{\frac{{\rm d}}{{\rm d}t}}\,\omega (t)+{\frac{\hbar \lambda }{m}}\,=i[\hbar
\kappa /m+\omega ^{2}(t)]\,,  \label{eq:26}
\end{equation}
with $\omega (0)=-{\frac{\hbar }{2m\sigma _{q}^{2}}}$, which define the
solution of (\ref{eq:21}) in the form given by (\ref{eq:23}). From (\ref%
{eq:24}) we get $\hat{q}(t)=-\Re \hat{w}(t)/2\Re \omega (t)$ which is the
root of the equation $R^{\prime }(t,x)=mU(t,x)=0$ for which the maximum of
the posterior density $|\hat{\varphi}(t,x)|^{2}=\exp [(2/\hbar )R(t,x)]$ is
attained. The posterior momentum $\hat{p}$ coincides with $mV(t,\hat{q}%
)=S^{\prime }(t,x)_{x=\hat{q}}$ and by (\ref{eq:24}) $\hat{p}(t)=m\Im
\lbrack \hat{w}(t)+\omega (t)\hat{q}(t)]$.

Eqs. (\ref{eq:24})--(\ref{eq:26}) give the Hamilton-Langevin equations
describing the time-development of posterior mean values of position and
momentum, 
\begin{equation}
\begin{array}{ll}
\hat p{\rm d} t-m{\rm d}\hat q=\hbar(\lambda/2)^{1/2} {\frac{{\normalsize 
{\rm d}\tilde Y}}{{\normalsize \Re\,\omega}}}\,, & \quad\hat q(0)=q\,, \\%
[+10pt] 
{\rm d}\hat p-mg{\rm d} t=\hbar\Big(\kappa\hat q{\rm d} t-(\lambda/2)^{1/2} {%
\frac{{\normalsize \Im\,\omega}}{{\normalsize \Re\,\omega}}}\,{\rm d}\tilde Y%
\Big)\,, & \quad\hat p(0)=p\,.%
\end{array}
\label{eq:27}
\end{equation}
The posterior position and momentum dispersions for the posterior wave
function in the form (\ref{eq:16}) are given by the formulas 
\begin{equation}
\tau^2_q(t)=-\,{\frac{\hbar}{2m\Re\omega(t)}}\,,\quad \tau^2_p(t)=-\,{\frac{%
\hbar m|\omega(t)|^2}{2\Re\omega(t)}}\,,  \label{eq:28}
\end{equation}
with $\omega(t)$ being the solution of eq. (\ref{eq:26}); therefore the
Heisenberg inequality $\tau^2_q(t)\tau^2_p(t)\geq\hbar^2/4$ is fulfilled.

The general solution of eq. (\ref{eq:26}) reads 
\begin{equation}
\omega(t)=i\alpha{\frac{\omega(0)+i\alpha\tanh(\alpha t)}{%
i\alpha+\omega(0)\tanh(\alpha t)}}\,,\quad \alpha=(\hbar/m)^{1/2}
(\kappa+i\lambda)^{1/2}\,.  \label{eq:29}
\end{equation}
Obviously, $\lim_{t\rightarrow\infty}\omega(t)=i\alpha$, i.e. $i\alpha$ is
the asymptotic stationary solution of (\ref{eq:26}). Consequently, posterior
dispersions of position and momentum tend to finite limits independent of
their initial values, from (\ref{eq:28}) and (\ref{eq:29}) we get 
\begin{eqnarray}
\tau^2_q(\infty)&=&{\frac{\hbar}{2m\Im\,\alpha}}= \left({\frac{\hbar}{2m}}%
\right)^{1/2}[(\kappa^2+\lambda^2)^{1/2}- \kappa]^{-1/2}\,,  \nonumber \\
\tau^2_p(\infty)&=&{\frac{\hbar m|\alpha|^2}{2\Im\,\alpha}}= \left({\frac{%
\hbar^3m}{2}}\right)^{1/2}\left({\frac{\kappa^2+\lambda^2}{%
(\kappa^2+\lambda^2)^{1/2} - \kappa}}\right)^{1/2}\,.  \label{eq:30}
\end{eqnarray}

Let us pay attention to the particular watchdog effects which can be
obtained from (\ref{eq:30}).

\begin{itemize}
\item[(a)] For $\kappa <0$, i.e. for the harmonic oscillator we find that
the width of the stationary Gaussian packet is smaller than that for the
unobserved oscillator. If the accuracy $\lambda \rightarrow \infty $ then $%
\tau _{q}^{2}(\infty )\rightarrow 0$ but $\tau _{p}^{2}(\infty )\rightarrow
\infty $.

\item[(b)] For $\kappa=0$, i.e. for a particle in a homogeneous (gravitation
or electric) field, in particular for a free observed particle ($g=0$) the
Gaussian packet does not spread out in time. The asymptotic localization $%
\tau^2_q(\infty)=(\hbar/2m\lambda)^{1/2}$ is inversely proportional to the
mass of the particle and the measurement accuracy coefficient.

\item[(c)] For $\kappa>0$, i.e. for the case of a linear active system
(harmonic accelerator) we obtain a similar watchdog effect as in case (b).
\end{itemize}

Thus the collapse problem of the wave function for a quantum particle under
the position measurement has found the dynamical solution.

\bibliographystyle{plain}
%\bibliography{belavkin,belbib,bestim,boson,grand}
%

\end{document}